# Self-assembly of well-separated AlN nanowires directly on sputtered metallic TiN films

M. Azadmand[1, 2, a), b)], T. Auzelle[1, a)], J. Lähnemann[1], G. Gao[1, c)], L. Nicolai[1], M. Ramsteiner[1], A. Trampert[1], S. Sanguinetti[2], O. Brandt[1], and L. Geelhaar[1]

[1]*Paul-Drude-Institut für Festköperelektronik, Leibniz-Institut im Forschungsverbund Berlin e.V., Hausvogteiplatz 5-7, 10117 Berlin, Germany*
[2]*L-NESS and Dipartimento di Scienza dei Materiali, Università di Milano - Bicocca, Via R. Cozzi 55, 20125 Milano, Italy*

[a)]*These authors contributed equally.*
[b)]*Email: azadmand@pdi-berlin.de*
[c)]*Present address: Material Science and NanoEngineering department, Rice University, 6100 Main Street, Houston, TX 77005, USA*

## ABSTRACT

We demonstrate the self-assembled formation of AlN nanowires by molecular beam epitaxy on sputtered TiN films on sapphire. This choice of substrate allows growth at an exceptionally high temperature of 1180 °C. In contrast to previous reports, the nanowires are well separated and do not suffer from pronounced coalescence. This achievement is explained by sufficient Al adatom diffusion on the substrate and the nanowire sidewalls. The high crystalline quality of the nanowires is evidenced by the observation of near band edge emission in the cathodoluminescence spectrum. The key factor for the low nanowire coalescence is the TiN film, which spectroscopic ellipsometry and Raman spectroscopy indicate to be stoichiometric. Its metallic nature will be beneficial for optoelectronic devices employing these nanowires as the basis for (Al,Ga)N/AlN heterostructures emitting in the deep ultraviolet spectral range.

## MAIN TEXT

AlN is the material of choice for optoelectronic devices working in the deep ultraviolet (DUV) spectral range.[1,2] For the growth of AlN epilayers, highly lattice-mismatched substrates such as sapphire or SiC are commonly employed. The substantial structural difference between substrate and epilayer leads to the formation of threading dislocations in high densities,[1] which limit device performance. Bulk AlN substrates have become available in the last few years, but are still prohibitively expensive. An attractive alternative is the synthesis of AlN in the shape of nanowires (NWs). The NW geometry enables the growth of AlN in high crystalline perfection on dissimilar substrates, as mismatch-induced strain can elastically relax at the free sidewall surfaces[3] of NWs with sufficiently small diameters, while in thicker NWs dislocation lines are confined to the interface region and do not propagate into the upper part of the NW.[4,5] The synthesis of AlN NWs has been explored by various techniques[6] but band edge luminescence from self-assembled AlN NWs has been reported exclusively for growth by molecular beam epitaxy (MBE).[7,8] The self-assembled formation of AlN NWs is very challenging, and efforts by MBE can be categorized into three different approaches by the type of substrate: i) Si(001) with a thin SiO$_2$ layer,[7] ii) Si(111) with a thin Si$_3$N$_4$ layer,[9] and iii) short GaN NW stems on Si(111).[8] All these approaches result in NW ensembles that are highly coalesced.

Pronounced coalescence is also typical for the self-assembled growth of GaN NWs by MBE, usually on Si substrates, and has been studied in depth for this material system.[10] Coalescence leads to the formation of dislocations, thus imposing limitations on crystalline perfection, and is mainly caused by bundling of close-by NWs. Recently, we achieved the growth of uncoalesced GaN NW ensembles on crystalline TiN films, which was enabled by a sufficiently low NW number density resulting from diffusional repulsion of nuclei on the substrate surface.[11] Inspired by this finding, we explore in this study the growth of AlN NWs on TiN.

The key advantage of TiN in this context is the combination of its high chemical and thermal stability. In particular, TiN does not react with AlN or its constituent elements. This property facilitates the use of substrate temperatures above 1000 °C to promote Al adatom diffusion, both on the substrate to reduce the NW density and on the NW sidewalls to avoid lateral growth, in order to obtain well separated NWs. An additional advantage of TiN is its excellent electrical and thermal conductivity, which enhances charge injection and waste heat dissipation for devices. In our previous GaN NW studies, we synthesized the TiN films by sputtering Ti onto sapphire substrates followed by nitridation under the N plasma in the MBE chamber. For this approach, interfacial reactions during nitride NW growth were hard to avoid.[12] In order to start with a chemically stable substrate surface, we now employ reactive sputtering[13] to fabricate stoichiometric crystalline TiN films prior to the MBE process.

TiN films with a thickness of 0.75 μm were deposited on 2-inch Al$_2$O$_3$(0001) substrates using a magnetron sputtering system with a Ti target. The pressure was $10^{-3}$ mbar, the Ar flow rate 12±0.2 sccm, the N$_2$ flow rate 3±0.2 sccm, the DC plasma power about 500 W, and the sputter duration 40 min. During the sputtering, a bias of 100 V was applied to the substrate to improve the physical properties of the TiN film.[14] Without exposure to air, the TiN films on sapphire were transferred to the MBE growth chamber. There, the substrate temperature was raised to 930 °C, and then increased further to 1180 °C over the course of 25 min while the TiN was exposed to an active N flux provided by an RF plasma source operated at 500 W and 2.5 sccm N$_2$ flow. The substrate temperature was measured by a pyrometer, taking into account the emissivity of TiN (0.23 at this temperature) and absorption in the viewport. Reflection high-energy electron diffraction (RHEED) revealed the pattern expected for TiN, as observed in Ref. [15]. The growth of AlN was initiated by providing an Al flux of $1.7 \times 10^{14}$ atoms cm$^{-2}$s$^{-1}$ (0.14 ML/s) supplied by a conventional Knudsen cell, whereas the active N flux was not changed and corresponded to $1.03 \times 10^{15}$ atoms cm$^{-2}$s$^{-1}$ (1.03 ML/s). The resulting N-rich conditions are typically also chosen for the self-assembly of GaN NW ensembles. The growth duration was 270 min. RHEED showed the first AlN-related diffraction spots 20 min after opening the Al shutter. This delay evidences that nucleation of AlN on sputtered TiN is preceded by an incubation time, similarly to what is characteristic for GaN NW growth.[16]

A TiN film of a reference sample without AlN growth was analyzed by variable-angle spectroscopic ellipsometry at room temperature and van der Pauw measurements. The morphology of the AlN NW samples was investigated using a scanning electron microscope (SEM). In addition, the microstructure of dispersed NWs was studied by high-resolution transmission electron microscopy (HRTEM) at 200 kV. To determine the lattice constant as well as inhomogeneous strain, tilt, and twist of as-grown NWs, we employed X-ray diffractometry (XRD) using Cu Kα1 radiation, a Ge(220) hybrid monochromator, and a Ge(220) analyzer. Furthermore, both the AlN NWs and the TiN film were studied at room temperature by Raman spectroscopy utilizing a laser wavelength of 473 nm. To examine the optical emission of the AlN NWs, we employed cathodoluminescence (CL) spectroscopy at low temperature (10 K) in a field-emission SEM operated at 5 kV and equipped with a high sensitivity photomultiplier tube as detector. The spectra were corrected for the instrument response.

Bird's-eye and top view SEM images of the AlN NWs are shown in Fig. 1(a) and 1(b), respectively. The NWs grow vertically with a number density of about $1\times10^{10}$ cm$^{-2}$ and a fill factor of 35%. They exhibit, on average, a height of 850 nm and a diameter of 55 nm. The former value corresponds to an average vertical growth rate of 3.4 nm/min (taking into account the incubation time), i.e., much higher than the 2.1 nm/min expected for planar AlN growth under these fluxes. This discrepancy implies that, in addition to the direct impingement of atoms on the NW top facet, there is a significant contribution of adatom diffusion on the sidewalls towards the tip. This phenomenon was previously observed and explained for the growth of GaN NWs,[17] but has not been reported for AlN NWs. In our case, the extremely high growth temperature promotes adatom diffusion, thus enhancing vertical growth and reducing parasitic lateral growth.

The key advancement with respect to morphology is that our AlN NWs grow well separated from each other. Following the method proposed by Brandt et al.,[18] we extract a coalescence degree of 47% from the circularity of the NW top facets (for a circularity threshold of 0.762). In comparison, typical GaN NWs grown on Si exhibit coalescence degrees of 70% and higher,[18] and only in an exceptional case 49%,[19] while GaN NW ensembles on TiN are essentially uncoalesced (7%).[11] The decisive factor for the low coalescence degree of GaN NWs on TiN is substantial Ga adatom diffusion on the substrate surface during the nucleation phase, which leads to a fairly low NW number density of around $1\times10^9$ cm$^{-2}$. The number density of AlN NWs is about an order of magnitude higher, which implies that the Al adatom diffusion length on the substrate surface is smaller than the one of Ga (at around 400 °C lower substrate temperature). The absolute number density we observe here is similar to usual values for GaN NWs on Si. In the latter case, coalescence has been found to result from the bundling of neighboring NWs, where above a characteristic height the gain in surface energy outweighs the elastic energy of bending.[10] Radial growth was found to be irrelevant after NW formation. For the number density and diameter of our AlN NWs, the characteristic height for bundling is 470 nm. Since the NWs are much longer, one would expect pronounced bundling. However, Fig. 1(a) does not reveal any bundling events, and this finding is confirmed by dedicated cross-sectional micrographs (see supporting information). Possibly, bundling does not occur because it is kinetically inhibited. Furthermore, the substantial adatom diffusion on the sidewalls evidenced in the current study suggests that, under our conditions, also for AlN NWs radial growth does not contribute significantly to coalescence. Thus, a detailed analysis of coalescence in such AlN NW ensembles would require further experiments.

TEM images of a representative AlN NW are depicted in Fig. 1(c) and 1(d). The NW is delimited by well-defined top and sidewall facets, exhibits the wurtzite structure, and is free of extended defects. Further TEM measurements revealed that the interface with the TiN substrate layer is rough and that the interface region is defective (not shown). The NW axis corresponds to the ⟨0001⟩ axis of the wurtzite crystal, and we identified the crystallographic orientation of the sidewall facets following the method reported in Ref. [20], by combining top-view SEM images with X-ray pole figures. Remarkably, $\{11\bar{2}0\}$ sidewalls are found, although according to density-functional theory calculations the surface energy of these planes is larger than for {1$\bar{1}$00} facets.[21] The latter ones form on self-assembled GaN NWs, but {11$\bar{2}$0} sidewalls were reported also for (Al,Ga)N NWs grown by MBE.[22] Our result suggests that the growth of AlN NWs is strongly influenced by kinetics. The high-resolution image in Fig. 1(d) shows that the top facets enclose an angle of 30±1° with the (0001) plane. Hence, the top facets can belong to the {11$\bar{2}$5} or {1$\bar{1}$03} family of planes, with the respective angles being 29° and 31.64°. Independent of the precise orientation, the top facets of these AlN NWs are systematically formed by semi-polar planes, although the pyramidal shape of the tip is not always as symmetric as in Fig. 1(c). An important parameter of nitride crystals is their polarity. As detailed in the supporting information, etching experiments that we carried out for our AlN NWs are inconclusive, while high-resolution high-angle annular dark-field micrographs acquired for one dispersed NW show that it is Al-polar, i.e. growth takes place in the [0001] direction (see supporting information). This result is confirmed by piezoresponse force microscopy studies,[23] but it is surprising given that self-assembled GaN NWs grow along the N-polar [000$\bar{1}$] direction.[24,25]

The key factor that enables the growth of these NWs is the TiN substrate. The sputtered TiN films exhibit a golden color, which is a first indication that the TiN is stoichiometric.[26] Variable-angle spectroscopic ellipsometry was employed for a more detailed analysis. Figure 2(a) and 2(b) show the real part ⟨$\varepsilon_1$⟩ and imaginary part ⟨$\varepsilon_2$⟩ of the measured pseudo dielectric function, respectively, along with reference data from literature.[27] The behavior of ⟨$\varepsilon_1$⟩ and ⟨$\varepsilon_2$⟩ at low photon energies is characteristic for a metal. The plasma energy extracted from the zero crossing of ⟨$\varepsilon_1$⟩ (≈ 2.6 eV) corresponds to stoichiometric TiN.[27,28] Also, the resistivity of our sputtered TiN films was measured to be as low as 3×10$^{-4}$ Ωcm, which is consistent with the metallic nature and previously reported values for TiN.[14] Figure 2(c) presents the Raman spectrum of an as-grown sample with spectral features of both the TiN film and the AlN NWs. The position of the low frequency transverse acoustic phonon mode of TiN is 200 cm$^{-1}$, in agreement with the value for stoichiometric TiN in Ref. [29]. Thus, multiple techniques confirm that our reactively sputtered TiN films are stoichiometric. Furthermore, the XRD profile shown in Fig. 2(d) does not contain any reflections from TiN$_x$O$_y$ alloys, demonstrating that even at the very high substrate temperature used for AlN growth, the TiN film does not react with the Al$_2$O$_3$ substrate.

Raman spectroscopy allows us in addition to assess the strain of the AlN NWs. As depicted in the inset to Fig. 2(c), the position of the E$_2^H$ optical phonon line corresponding to high-frequency optical phonons is almost identical for the NWs and bulk AlN measured as a reference side by side. This agreement suggests that the NWs are essentially free of strain, as expected for the NW geometry. For a more precise measurement, we extracted from ω–2θ XRD scans as seen in Fig. 2(d) the angular position of the 0002 and 0004 Bragg reflections for the AlN NWs. From a fit to the two positions, we determined a value of 0.49835 ± 0.00005 nm for the c lattice constant. This value is higher than the recently reported value (0.49808 nm) for bulk (freestanding) AlN[30] and corresponds to an out-of-plane strain $\epsilon_{zz}$ = 5×10$^{-4}$. If this strain were hydrostatic, it would result in a clearly detectable shift in the Raman mode. Hence, we attribute the out-of-plane strain to surface stress acting along the NW axis. Some of the present authors recently found the same phenomenon in a detailed XRD study of similarly sized GaN NWs.[31] In sufficiently thin NWs, surface stress can be significant, amounting to, e.g., $\epsilon_{zz}$ = 2.4×10$^{-4}$ for GaN at an average diameter of 58 nm. Also, the effect of the out-of-plane strain component on the frequency of phonon modes and the energy of exciton transitions can be partly or even completely compensated by the in-plane strain component.

Furthermore, we extracted the angular width of two Bragg reflections for both the AlN NWs and the $Al_2O_3$ substrate and used these values for a Williamson-Hall plot (cf. supporting information). From the slope of this plot, we deduced a root mean square (rms) strain variation in the AlN NWs of $2\times10^{-4}$. This rms strain is typically interpreted as inhomogeneous strain. A similar value ($1.5\times10^{-4}$) was observed for GaN NW ensembles on Si with comparable coalescence degree (49%), and was associated with coalescence and the mutual misorientation of NWs.[19] However, in view of the importance of surface stress in our AlN NWs, its dependence on the NW diameter in combination with the distribution of the latter may also contribute.[31] The average AlN NW tilt and twist was determined from XRD rocking curves around the AlN(0002) and AlN($1\bar{1}00$) reflections to be about 1.6 and 2.1°, respectively, again similar to values for GaN NWs on Si.[19]

Figure 3 presents a low-temperature CL spectrum of the AlN NWs in comparison to a reference spectrum of free-standing bulk AlN featuring narrow excitonic lines with well-known energies.[32] The inset shows that both samples exhibit an intense emission line close to the band edge at 6 eV and the commonly observed emission from deep levels at 3.9 eV. The latter has been attributed to complexes of O with Al vacancies.[33] As seen in the main diagram, the near-band-edge emission of both samples peaks at the same energy, consistent with the Raman results. For the bulk sample, it is known that the dominant transitions are the donor-bound exciton ($D^0,X_A$) at 6.012 eV and the free A exciton ($X_A$) at 6.034 eV. The emission band of the AlN NWs is broader and displays a shoulder at low energy. The width cannot be explained by the rms strain variation determined by XRD. Instead, we attribute the broadening to potential fluctuations caused by a random distribution of ionized donors,[34] possibly related to O evaporation from uncovered parts of the sapphire wafer. The low-energy shoulder is too strong to be related to a phonon replica, and might be related to stacking faults in the defective region close to the substrate.

In summary, employing stoichiometric, reactively-sputtered TiN films on sapphire as a substrate enables the growth of well-separated AlN NWs at an exceptionally high temperature of 1180 °C. For this approach, starting NW growth with GaN stems can be avoided. These AlN NWs have a high degree of crystalline perfection and exhibit near-band-edge luminescence at the same position as for the bulk AlN reference. This work opens up the possibility to use AlN NW ensembles as quasi-substrate for the growth of high quality AlN/(Al,Ga)N heterostructures and the realization of devices operating in the deep UV range that would also benefit from the metallic nature of TiN.

We are indebted to M. Bickermann, Leibniz-Institut für Kristallzüchtung, Berlin, for providing the bulk AlN reference sample. We are grateful to K. Morgenroth for her support during sample preparation as well as for maintenance of the MBE system together with C. Stemmler and M. Höricke. We thank A.-K. Bluhm for SEM characterization and M. Heilmann for a critical reading of the manuscript. Funding from the German Bundesministerium für Bildung und Forschung through project No. FKZ:13N13662, from Regione Lombardia (Italy) through project COSIMTO, and from the European Regional Development Fund (ERDF) through project No. 2016011843 is gratefully acknowledged.

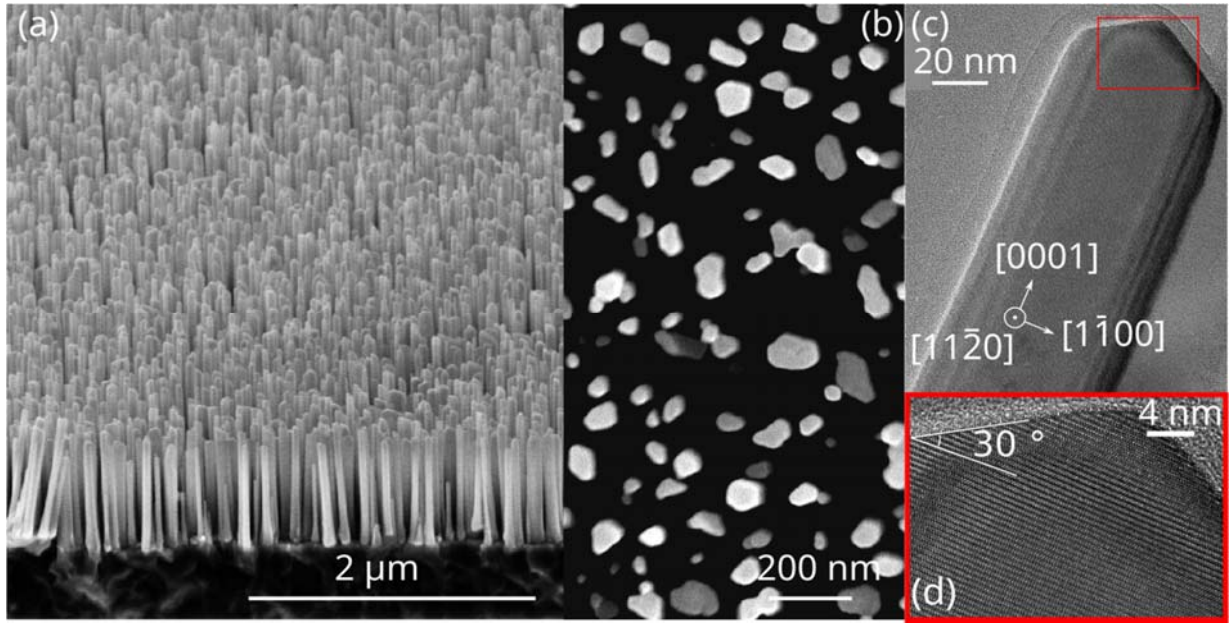

Fig. 1. (a) Bird's eye and (b) top view SEM images of AlN NWs grown on a reactively-sputtered TiN film. (c) Cross-sectional bright field and (d) high-resolution TEM images from the tip of a dispersed AlN NW.

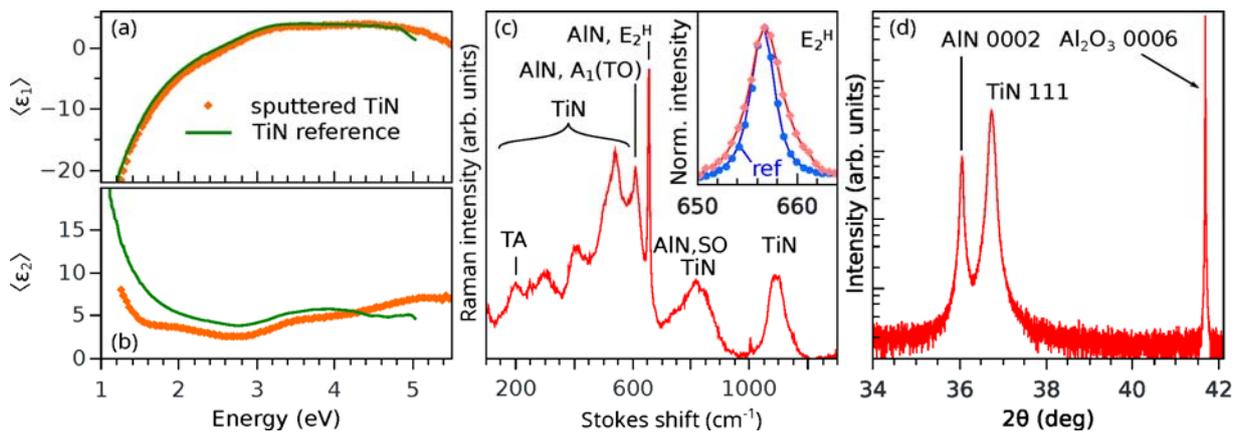

Fig. 2. (a) Real and (b) imaginary parts of the pseudo dielectric function of a sputtered TiN film (orange symbols), obtained by spectroscopic ellipsometry. The reference TiN spectrum (green line) is taken from Ref. [27]. (c) Raman spectrum and (d) ω–2θ XRD scan of the sample with AlN NWs. The inset to (c) displays a magnification of the $E_2^H$ phonon line, in direct comparison to a reference (ref) sample of bulk AlN (blue).

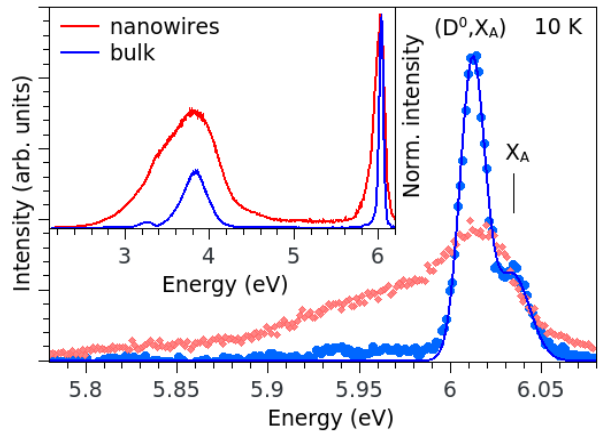

Fig. 3. Low temperature CL spectra of the AlN NW ensemble grown on sputtered TiN (red diamonds) and of freestanding bulk AlN (blue spheres). The line is a Gaussian lineshape fit for the bulk sample, taking into account the $(D^0,X_A)$ and $(X_A)$ transitions. The inset shows the data over an extended energy range, normalized to the near-band-edge emission.